\documentclass[floatfix, 
reprint, 
amsmath, 
amssymb, 
aps, 
pra]{revtex4-2}

\usepackage{graphicx}
\usepackage{dcolumn}
\usepackage{bm}
\usepackage{xcolor}
\usepackage{braket}
\usepackage[normalem]{ulem}

\begin{document}

\preprint{APS/123-QED}

\title{Decoherence of surface phonons in a quantum acoustic system}

\author{Camryn Undershute}
\email{undershu@msu.edu}
\author{Joseph M. Kitzman}%
\author{Camille A. Mikolas}
\author{Johannes Pollanen}
\email{pollanen@msu.edu}
\affiliation{%
 Department of Physics and Astronomy, Michigan State University, East Lansing MI 48824 USA\\
}%

\date{\today}

\begin{abstract}
Phononic resonators are becoming increasingly important in quantum information science, both for applications in quantum computing, communication and sensing, as well as in experiments investigating fundamental physics. Here, we study the decoherence of phonons confined in a surface acoustic wave (SAW) resonator strongly coupled ($g/2\pi\approx 9$ MHz) to a superconducting transmon qubit. By comparing experimental data with numerical solutions to the Markovian master equation, we report a surface phononic energy decay rate of $\kappa_1/2\pi=480$ kHz and a pure dephasing rate of $\kappa_{\phi}/2\pi=180$ kHz. These rates are in good agreement with the level of decoherence we extract from qubit-assisted spectroscopic measurements of the SAW resonator. We additionally find that the timescales over which coherent driven dynamics and decoherence occur are comparable, highlighting the need to model the composite device as an open quantum system. We discuss possible sources of decoherence in SAW-based quantum acoustic devices and the application of these devices in future quantum acoustic dissipation engineering experiments. The decoherence characterization techniques we employ are broadly applicable for investigating and benchmarking the effects of loss and dephasing on mechanical resonators in the quantum regime.
\end{abstract}

\maketitle

\section{\label{sec:intro}Introduction}

Engineered coupling in hybrid quantum systems enables the manipulation and measurement of composite quantum systems having disparate, but complementary, properties~\cite{Clerk2020hybrid}. This is particularly true of mechanical and phononic degrees of freedom, which can be integrated with a wide variety of quantum systems including defect centers in diamond~\cite{macquarrie2013,golter2016coupling,Maity2020}, quantum dots~\cite{mcneil2011demand,decrescent2022large}, trapped electrons~\cite{Byeon2021} and superconducting qubits~\cite{lahaye2009nanomechanical, chu2017quantum,satzinger2018quantum}. The last of these examples has led to the development of a framework known as circuit quantum acoustodynamics (cQAD), in which superconducting qubits interact with microwave frequency mechanical resonators in a fashion analogous to their interaction with microwave frequency photons, i.e. circuit quantum electrodynamics (cQED). Recent results have demonstrated the viability of cQAD for applications in on-chip information transfer~\cite{gustafsson2014propagating, dumur2021quantum}, generating remote entanglement between distant qubits~\cite{bienfait2019phonon, dumur2021quantum}, developing quantum memories~\cite{hann2019hardware}, and creating novel methods for quantum computation using mechanical degrees of freedom~\cite{bild2023schrodinger, qiao2023splitting, von2023engineering, yang2024mechanicalqubit}. Beyond applications for quantum information processing, phononic resonators are used in numerous experiments investigating open questions in fundamental physics. Acoustic resonators have been used to probe novel regimes of quantum optics~\cite{guo2017giant, andersson2019non,BUTTTS2023superconductingsingleatomphononlaser}, perform macroscopic tests of quantum mechanics~\cite{bild2023schrodinger, schrinski2023macroscopic}, implement delayed-choice quantum erasure~\cite{bienfait2020quantum} and search for high-frequency gravity waves~\cite{goryachev2021rare}.

While tremendous progress has been made integrating mechanical resonators with superconducting qubit-based hybrid systems, phononic degrees of freedom can also introduce new channels for dissipation and decoherence. When this dissipation is uncontrolled, it is detrimental to the performance of quantum devices. However, quantum systems subject to controlled dissipation have been shown to exhibit a range of interesting phenomena. Dissipation in quantum systems can be used to simulate non-Hermitian Hamiltonians~\cite{chen2021quantum}, prepare and stabilize highly non-classical states~\cite{shankar2013autonomously} and even assist in quantum error correction~\cite{kapit2017upside, touzard2018coherent}. In these examples of reservoir engineering protocols, dissipation must be well understood so that it can be harnessed as an experimental resource. Most of these types of experiments have been demonstrated in cQED systems, where the controlled loss channels are photonic in nature. However, superconducting qubits interact not only with photonic loss mechanisms, but also with the ubiquitous bath of phonons~\cite{wilen2021correlated, chen2023phonon, kono2024mechanically, zhou2024observation}, opening the door for a class of novel phononic reservoir engineering experiments~\cite{kitzman2023bath}.

Understanding decoherence is also essential for advancing cQAD based technologies for applications in quantum information science. Consequently, experiments have investigated the decoherence of quantum states hosted in phononic crystals~\cite{wollack2022quantum, cleland2023studying}, bulk acoustic wave resonators~\cite{chu2017quantum, chu2018creation, chou2020measurements} and surface acoustic wave (SAW) devices~\cite{manenti2016surface, satzinger2018quantum}. The decoherence of SAW resonators is of particular interest, as the underlying mechanisms for phonon loss lack a comprehensive microscopic description~\cite{gruenke2024surface, tubsrinuan2024correlated}. Furthermore, many experiments reporting on the decoherence of multi-phonon states hosted in SAW resonators do so by performing only spectroscopic measurements. In this situation it is often difficult to distinguish the effects of interference and decoherence~\cite{rieger2023fano,kitzman2023fano}, in which case comparison against time-domain measurements would enable a more robust analysis of the scattering data.

In this work, we report on measurements and modelling of a cQAD system composed of a superconducting transmon qubit coupled to a SAW resonator. We measure both mechanical energy decay and dephasing of surface phonon coherent states hosted in the resonator using both time-resolved and spectroscopic techniques. From the pulsed measurements, we find that the time scales associated with the coherent and lossy dynamics of the surface phonons are comparable. We compare these results to numerical solutions of the Markovian master equation, which show excellent agreement with the data, revealing the open quantum system nature of the qubit-surface phonon hybrid device.

\section{\label{sec:exp-setup}Experimental setup}
\begin{figure}[h]
    \centering
    \includegraphics[width=3.288in]{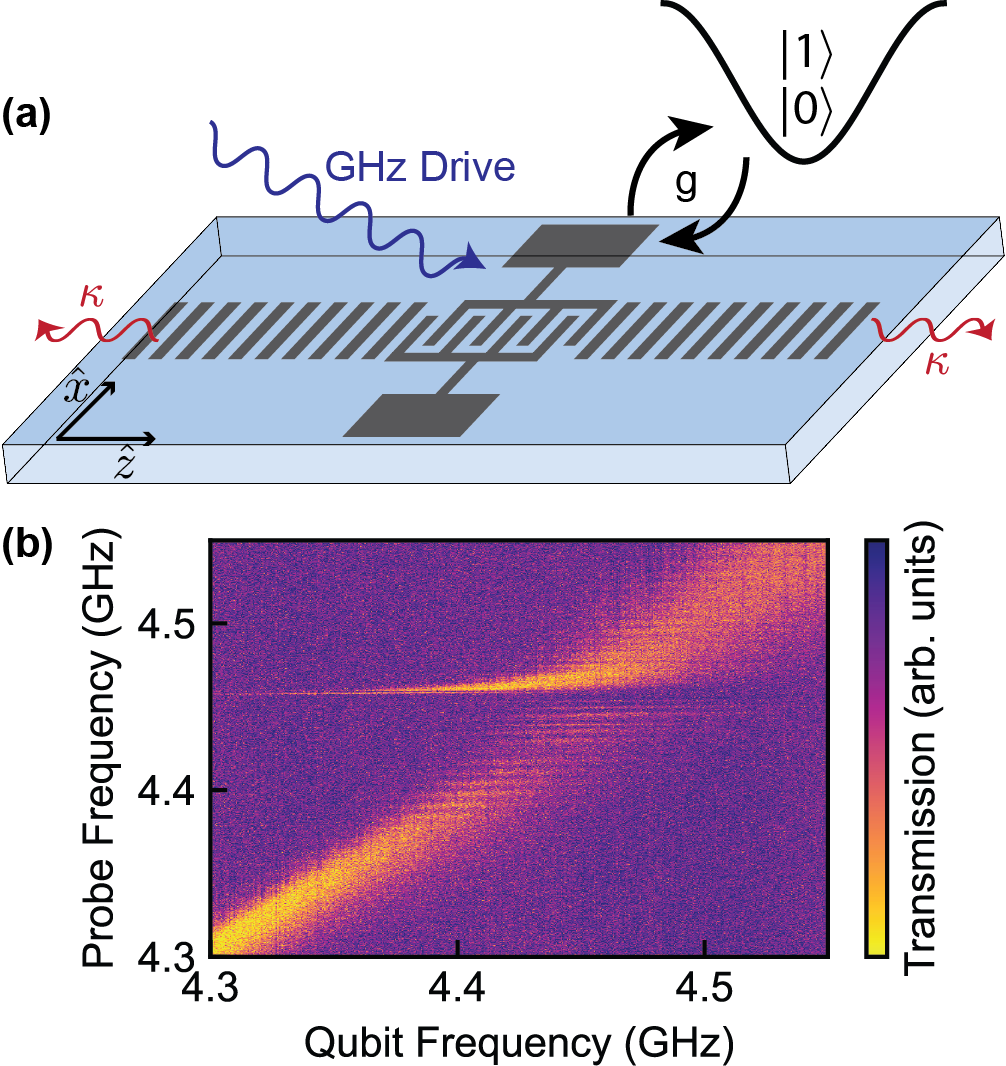}
    \caption{SAW-based quantum acoustics (a) Schematic of the experiment. The SAW device is composed of a metallic IDT and acoustic Bragg mirrors on a piezoelectric substrate (not drawn to scale). This device was designed to host a single resonant surface phonon mode. We drive this resonant SAW mode into a coherent state using an external classical microwave drive (dark blue), and then readout the mean phononic occupation via dispersive coupling to a transmon qubit. The SAW resonator undergoes decoherence at the rate $\kappa$. (b) SAW-qubit avoided crossing, from which we extract a mechanical coupling strength $g/2\pi=9$~MHz.}
    \label{fig:schematic}
\end{figure}
The experiment consists of a SAW resonator coupled to a transmon qubit. The SAW resonator is comprised of an inter-digitated transducer (IDT), which is responsible for transducing electromagnetic and mechanical energy, sandwiched between two acoustic Bragg mirrors, as depicted in Fig.~\ref{fig:schematic}(a). The geometry of the SAW device was designed to host a single resonant mode at $\omega_{m}/2\pi=4.45$~GHz using the coupling-of-modes numerical method~\cite{morgan2010surface}. The SAW resonator was fabricated using standard electron beam lithography techniques to pattern aluminum on piezo-electric \emph{Y}-cut LiNbO$_3$ (see~Ref.~\cite{kitzman2023bath} for fabrication details and design parameters). To manipulate and readout the state of the SAW resonator, we couple it to a flux tunable transmon qubit fabricated on a separate die. The transmon was fabricated by double-angle evaporation of aluminum on high-resistivity silicon. The qubit has a maximum Josephson tunnelling energy of $E_J/h=19.7$~GHz and a capacitive charging energy of $E_C/h=318$~MHz. At the frequency at which we operate the qubit for the measurements described below, $\omega_q/2\pi= 4.220$~GHz, it has a lifetime of $T_1=494$~ns and a decoherence time of $T_2=750$~ns, from which we calculate a pure dephasing time of $T_{\phi}=3.11$~$\mu$s.

The SAW device and qubit are assembled in a flip-chip configuration~\cite{kitzmanthesis} and placed in a three-dimensional (3D) electromagnetic cavity with frequency $\omega_c/2\pi~=~4.788$~GHz, which serves for qubit state readout. Two large (250~$\mu$m~$\times$~250~$\mu$m), metallic antenna pads have additionally been fabricated on both the SAW device and the qubit, which are galvanically connected to the IDT and SQUID loop, respectively. These pads facilitate the capacitive coupling between the SAW and qubit, as well as the coupling between each device and the electromagnetic field in the 3D cavity. The experiment is mounted on the base plate of a dilution refrigerator ($T\approx 15$~mK), at which temperature the SAW resonator, transmon and microwave cavity are all cooled near their quantum mechanical ground state, $\hbar\omega_{m},~\hbar\omega_q,~\hbar\omega_c~\gg~k_BT/\hbar$. All SAW, qubit and cavity readout and manipulation pulses were generated using a Xilinx RFSoC ZCU216 programmed with the recently developed supporting firmware~\cite{stefanazzi2022qick}, along with a single additional microwave source.

We confirm the coupling between the qubit and the SAW resonant mode by sweeping the flux that tunes the qubit frequency and measuring its absorption spectrum, as shown in Fig.~\ref{fig:schematic}(b). The qubit and resonant SAW mode undergo an avoided crossing, from which we extract a mechanical coupling strength $g/2\pi=9$~MHz \cite{schuster2007thesis}.

\section{\label{sec:techniquesmaybe}Measurement techniques and spectroscopy}
\subsection{\label{sec:popandro}Mechanical state population and readout}
To measure the phononic population of the SAW resonant mode, we flux-tune the qubit to $\omega_q/2\pi= 4.220$~GHz, which is well into the dispersive regime, $g/|\Delta|\ll 1$, with $\Delta = \omega_q-\omega_{m}$. In this regime, the hybrid SAW-qubit system obeys the Hamiltonian ($\hbar=1$),
\begin{equation}\label{eq:H_dispersive}
    \hat{H} = \omega_{m}(\hat{a}^{\dagger}\hat{a}+1/2)+(\omega_q+2\chi \hat{a}^{\dagger}\hat{a})\hat{\sigma}_z/2.
\end{equation}
Here, $\hat{a}^{\dagger}$ and $\hat{a}$ are the phononic creation and annihilation operators acting on the SAW subspace and $\hat{\sigma}_z$ is the Pauli operator acting on the qubit subspace. As seen from Eq.~\ref{eq:H_dispersive}, the effective frequency of the qubit, $\tilde{\omega}_q=\omega_q+2\chi\hat{a}^{\dagger}\hat{a}$, acquires a shift that depends on the population of the SAW resonant mode. The frequency shift per phonon, $2\chi/2\pi$, is given by~\cite{koch2007charge-insensitive},
\begin{equation}\label{eq:chi}
    2\chi = 2\frac{-g^2}{\Delta}\frac{\alpha}{\Delta-\alpha},
\end{equation}
where $\alpha/2\pi=318$~MHz is the anharmonicity of the transmon qubit, as Eq.~\ref{eq:chi} takes into account the multi-level nature of the transmon. At the operating qubit frequency, we calculate $2\chi/2\pi\simeq -0.4$~MHz. This resonator state-dependant qubit frequency shift, i.e. Stark shift, has been widely studied in both cQED, in which it underpins nondestructive qubit state readout~\cite{schuster2005ac, gambetta2006qubit}, and more recently in cQAD architectures~\cite{manenti2017circuit, arrangoiz2019resolving, sletten2019resolving,von2022parity, kitzman2023fano}. This Stark shift is often used as a tool, not only for sensing the bosonic population in the resonator, but also to facilitate logic gates in quantum computing protocols~\cite{chow2013microwave} or generate non-classical bosonic encodings by shuttling information from the qubit to the resonator mode~\cite{vlastakis2013deterministically}. As described below, we take advantage of the phononic Stark shift to perform qubit-assisted spectroscopy and time-resolved measurements of the SAW resonant mode population $\overline{n}$. 

\begin{figure}
    \centering
    \includegraphics[width=3.295in]{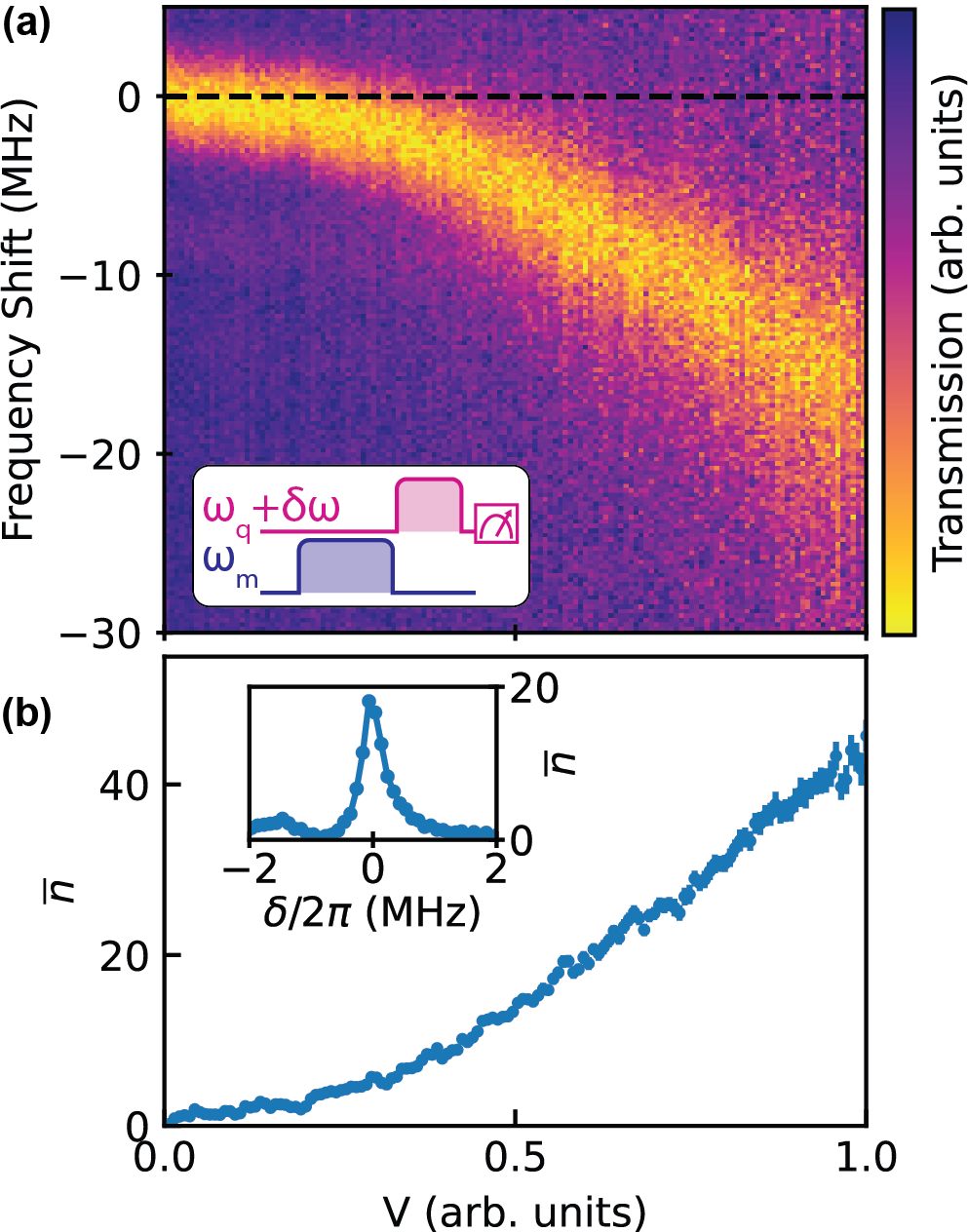}
    \caption{Population of the SAW coherent state via applied microwave drive. (a) The qubit absorption spectrum acquires an increasingly large shift from the bare qubit frequency $\omega_q$ (dashed black line) as we apply a drive with frequency $\omega_{m}$ and increasing drive amplitude $V$. In addition to the frequency shift, the qubit linewidth increases with $V$, which is further evidence that the SAW resonator is driven into a coherent state. Inset shows the pulse sequence used for this spectroscopic measurement. (b) Extracted mean phononic occupation $\overline{n}$ calculated from the qubit frequency shift. Inset shows the SAW resonator absorption spectrum. For fixed drive length ($t_{d}=10$~$\mu$s) and drive amplitude ($V~=~0.72$~arb.~units), we sweep the frequency of the SAW resonator population drive $\omega_{m}+\delta$ and measure $\overline{n}$.}
    \label{fig:sawdrive}
\end{figure}

To populate the SAW resonant mode we apply a coherent drive with frequency $\omega_{m}/2\pi$ to the 3D microwave cavity. This implies there is some SAW resonator-3D cavity coupling. However, we note that this coupling is small relative to the other coupling rates as we do not find evidence for it in other measurements. For instance, we do not measure a resolvable shift in $\omega_c$ when we drive the SAW resonator. 

Following the population drive, we measure the qubit spectrum to determine the phonon induced shift. As seen in Fig.~\ref{fig:sawdrive}(a), as the amplitude of this drive is increased, the qubit acquires an increasingly large Stark shift from $\omega_q$ (denoted by the black dashed line). While we do measure a well resolved shift in the dressed frequency of the qubit $\tilde{\omega}_q$, we do not see discrete peaks in the qubit spectrum, which would correspond to individual Fock components of the resonant SAW mode's state, an effect colloquially known as ``number splitting"~\cite{schuster2007resolving, arrangoiz2019resolving}. This result confirms that we are in the weak dispersive regime, i.e. the total qubit decoherence rate $\gamma_q$ or the total SAW decoherence rate $\kappa$ is larger than the dispersive shift ($2\chi<$~max\{$\gamma_q, \kappa$\}). As a result, we cannot resolve the distribution of the SAW resonator state, but we can nonetheless extract the mean phononic population $\overline{n}$ by fitting each qubit spectrum to a Lorentzian lineshape, and calculating $\overline{n}=\langle\hat{a}^{\dagger}\hat{a}\rangle = \frac{\tilde{\omega}_q-\omega_q}{2\chi}$, as shown in Fig.~\ref{fig:sawdrive}(b). Throughout the rest of the discussion, we rely on this technique to extract $\overline{n}$.

\subsection{\label{sec:3tone}Qubit-assisted SAW spectroscopy}
We can also leverage the qubit to measure the absorption spectrum of the SAW resonator. In this measurement, we fix the amplitude and length of the drive exciting the SAW phonons to perform a quasi-CW measurement ($t_{d}=10$~$\mu$s), and sweep the frequency of this population drive, $\omega_{d}~=~\omega_{m}+\delta$. As seen in the inset of Fig.~\ref{fig:sawdrive}(b), we find that the spectrum is peaked at $\omega_{m}/2\pi~=~4457.37$~MHz and we extract a linewidth (full-width at half-maximum) of $\kappa/2\pi~=~430 \pm 27$~kHz. Here the uncertainty is the standard deviation of repeated measurements at various $V$. Interestingly, we find no systematic dependence of $\kappa$ on $V$ in the experimentally accessible range of amplitudes investigated. This implies that the surface phonon decoherence in our device is not limited by an interaction with a bath of fluctuating two-level systems (TLSs)~\cite{muller2019towards, andersson2021acoustic}. We note that in addition to the main mode of interest, we see evidence of a second resonant mode roughly 2 MHz below $\omega_{m}/2\pi$, which is likely associated with a spurious SAW mode. This deviation in the SAW response from the one that was predicted by the coupling-of-modes numerical method may be due to a lack of precise knowledge of the material parameters at low temperatures, or difference between the modelled parameters and those that are actualized in fabrication.

\section{\label{sec:model}Modelling}

In order to extract Hamiltonian drive rates and SAW decoherence rates from the measured values of $\overline{n}$, we can relate the phononic occupation to the reduced SAW density matrix $\rho_{m}$ by tracing over the SAW excitation number $\overline{n}(t) = Tr(\hat{a}^{\dagger}\hat{a}\rho_{m}(t))$. The problem then becomes that of solving a master equation for the time evolution of $\rho_{m}$ ($\hbar=1$), 
\begin{equation}\label{eq:master_eq_2}
    \begin{split}
        \dot{\rho}_{m} = -i[H_{D}, \rho_{m}] &+\kappa_1\mathcal{D}[\hat{a}]\rho_{m} \\ &+\frac{\kappa_{\phi}}{2}\mathcal{D}[\hat{a}^{\dagger}\hat{a}]\rho_{m},
    \end{split}
\end{equation}
where $H_D$ is the Hamiltonian describing the coherent evolution of $\rho_{m}(t)$. The Linblad superoperator,
\begin{equation}
    \mathcal{D}[\hat{L}]\rho_{m} = \hat{L}\rho_{m}\hat{L}^{\dagger}-\{\hat{L}^{\dagger}\hat{L},\rho_{m}\}/2,
\end{equation}
describes the effect of the bath on $\rho_{m}(t)$. The second term in Eq.~\ref{eq:master_eq_2} describes energy decay as single phonon loss, which occurs at the rate $\kappa_1$, and the last term describes dephasing, which occurs at the rate $\kappa_{\phi}/2$. To model our data, we use the Qutip Python package~\cite{johansson2012qutip, johansson2013qutip2} to set up and numerically solve Eq.~\ref{eq:master_eq_2}, and then iteratively fit our data to these numerical solutions~\cite{newville2015}.

\begin{figure}
    \centering
    \includegraphics[width=3.375in]{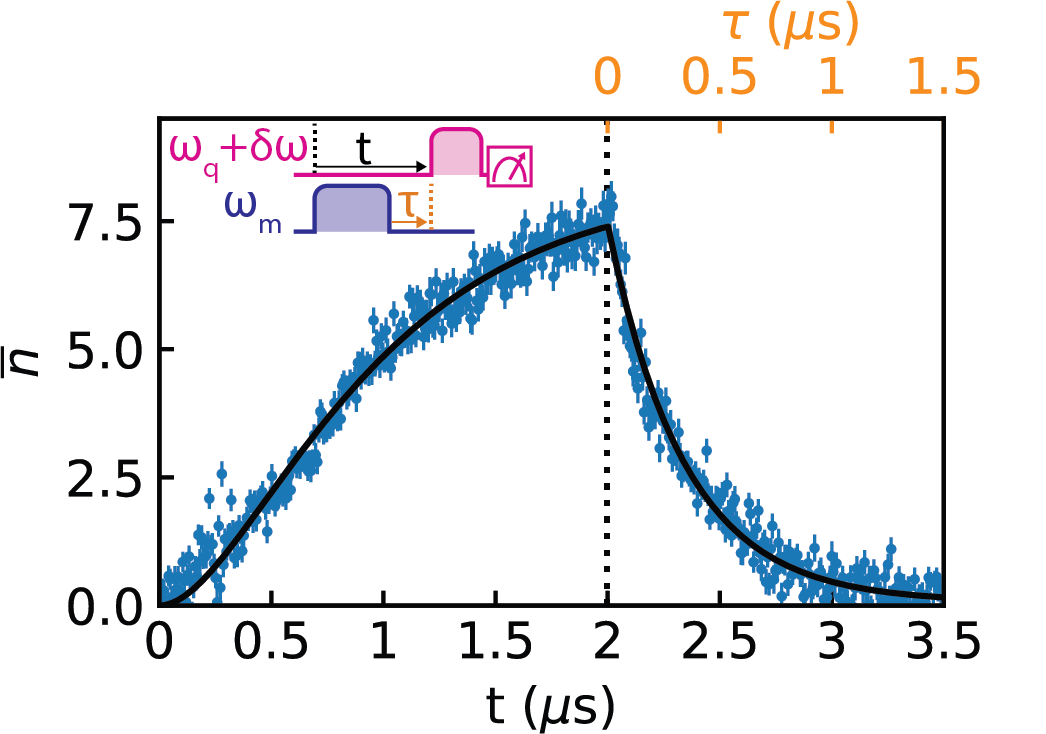}
    \caption{SAW coherent state ring-up-ring-down. The inset shows the pulse sequence for this measurement. For times $t<2$~$\mu$s, there is no delay between SAW drive and qubit readout ($\tau=0$) and the SAW is driven for the entire time $t$. For times $t>2$~$\mu$s, SAW drive length is fixed to 2~$\mu$s, denoted by the black dotted vertical line, and the SAW coherent state undergoes free evolution for time $\tau$ (indicated by the axis above the plot) before we readout its occupation via the qubit. Data and uncertainties are plotted in blue, and the black solid line is the fit to the model (see main text for details).}
    \label{fig:3}
\end{figure}
\section{\label{sec:results}Results and discussion}
\subsection{\label{sec:ringupringdown}Ring-up-ring-down}
To measure the energy decay of the SAW resonant mode, we fix the amplitude of the drive that populates the resonator $V$ and perform time-resolved measurements. As shown in Fig.~\ref{fig:3}, we measure the SAW coherent state ``ring-up", i.e. build up over time, by sweeping the length of the drive that excites the phononic population and then immediately reading out the resulting population via the qubit. Once the SAW resonant mode has been driven for $t_{d}=2$~$\mu$s we fix the drive length and allow the SAW to freely evolve for time $\tau$ before measuring the population, which decays over time, i.e. ``ring-down". Collectively, we refer to this series of experiments as a ``ring-up-ring-down" measurement. 

We fit these data to Eq.~\ref{eq:master_eq_2} as described in Sec.~\ref{sec:model}, where $H_{D}=U(t)(\hat{a}+\hat{a}^{\dagger})$ is in the frame rotating with $\omega_{m}$. Here, $U(t)$ is given by
\begin{equation}
    U(t)=\begin{cases}
        U, & \text{if }t<t_{d}\\
        0, & \text{otherwise}
    \end{cases},
\end{equation}
where $U$ is the driving rate associated with the drive amplitude $V$. In this fit, indicated by the solid black line in Fig.~\ref{fig:3}, $t_{d}=2$~$\mu$s and $\kappa_{\phi}=0$ are fixed so that the only free parameters to fit to are $U$ and $\kappa_1$. From this fit we get a driving rate of $U~=~4.35~\pm~0.03$~MHz for this drive amplitude and an energy decay rate of $\kappa_1/2\pi~=~480~\pm~30$~kHz.

There are multiple mechanisms that can contribute to loss of surface phonons from the resonator that likely play a role here. Similar to the Purcell effect in cQED systems, the spontaneous emission of the SAW resonator can be enhanced by its coupling to the qubit. However, we estimate that this contribution to the SAW decay rate, $(g/\Delta)^2(1/T_1)\approx 3.1$~kHz, is small. The Bragg mirrors that physically confine the SAW phonons have only a fractional reflectivity~\cite{morgan2010surface}. As a result, even phonons within the bandwidth of the mirrors have a small probability of leaving the device. Furthermore, in \emph{Y}-cut lithium niobate there is conversion from surface to bulk phonons~\cite{campbell1998surface} and the device architecture employed here is know to host parasitic bulk phonons modes~\cite{kitzman2023free}. Additionally, it is possible for surface phonon energy to be lost in conjunction with mass loading effects~\cite{morgan2010surface}. Finally, while our spectroscopic measurement implies that SAW lifetime is not dominated by an interaction with dielectric TLSs, nonetheless a fluctuating bath of weakly coupled TLSs could be contributing to $\kappa_{1}$~\cite{muller2019towards}. We note that while our analysis does not definitively identify which decay mechanism, or combination of mechanisms, dominate surface phonon loss in the device, the ``ring-up-ring-down" measurement protocol and analysis are a way to quantify loss, particularly when it is possible to systematically control various loss channels~\cite{gruenke2024surface}.

\subsection{\label{sec:ramsey}Mechanical dephasing}

\begin{figure}
    \centering
    \includegraphics{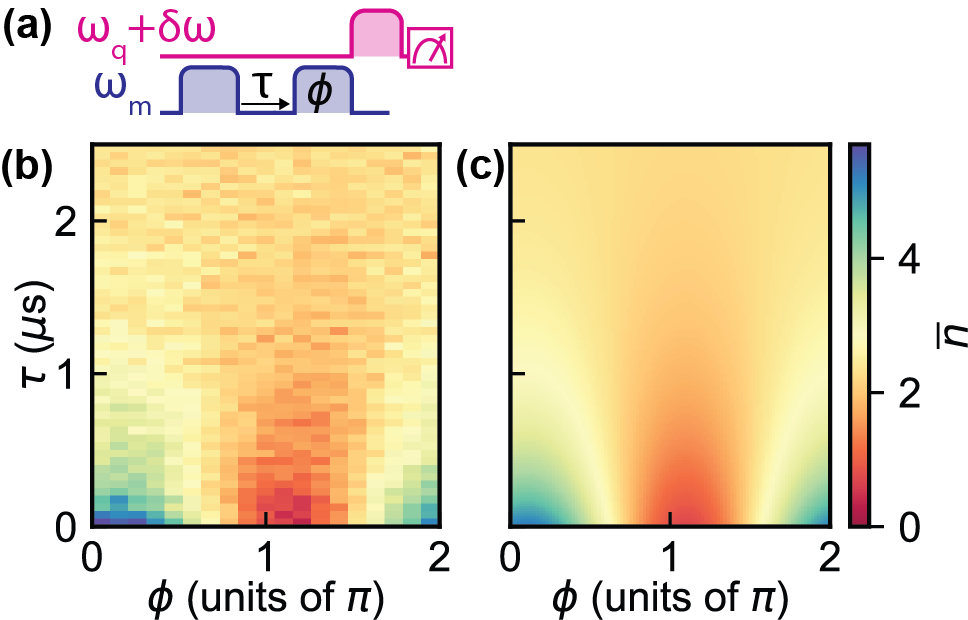}
    \caption{SAW dephasing. (a) SAW Ramsey pulse sequence. The first SAW population pulse is followed by a variable delay time $\tau$, and then a second SAW population pulse is applied with the same amplitude and duration, but a variable phase offset relative to the first SAW pulse. Subsequently we measure the mean phononic population via the qubit. (b) SAW Ramsey measurement results. For short $\tau$, the phononic occupation varies sinusoidal with $\phi$. As $\tau>$min\{$2\pi/\kappa_{\phi}, 2\pi/\kappa_{1}$\} the final phononic state size is independent of $\phi$. Data are fit to numerical solutions to Eq.~\ref{eq:master_eq_2}, and the resulting model is shown in (c).}
    \label{fig:Ramsey}
\end{figure}

In addition to measuring the energy decay of the SAW resonator, we also measure the mechanical dephasing by using a modified Ramsey sequence~\cite{cleland2023studying}. This consists of two Ramsey drives of fixed length $t_{d}~=~0.5$~$\mu$s and amplitude applied to the SAW resonator, immediately followed by SAW population readout, as seen in Fig.~\ref{fig:Ramsey}(a). To improve measurement contrast, in addition to varying the time between the two Ramsey drives $\tau$, we also vary the relative phase of the second drive $\phi$. 

The measured phononic population $\overline{n}(\tau, \phi)$ is shown in Fig.~\ref{fig:Ramsey}. We again fit a complete data set to numerical solutions to the master equation, where now the Hamiltonian accounts for both drives during the Ramsey pulse sequence: 
\vspace{-0.2cm}
\noindent \begin{equation} 
    H_D(t, \phi) = U(t, \phi)(\hat{a}^{\dagger}+\hat{a})+h.c., 
\end{equation}

\noindent where

\noindent \begin{equation}
    U(t, \phi)=
    \begin{cases}
        U, & \text{if }0<t<t_{d} \\
        U\cos(\phi), & \text{if }\tau<t<\tau+t_{d} \\
        0, & \text{otherwise.}
    \end{cases}
    \label{eq:ramsey_drive_amp}
\end{equation}

\noindent Now, in addition to fitting for $U$ and $\kappa_1$, we also leave $\kappa_{\phi}$ as a free parameter to be extracted from the fit. We note, the data is best fit when we include a phase offset $\phi_0$, by replacing $\phi$ with $\phi+\phi_0$ in Eq.~\ref{eq:ramsey_drive_amp} and leaving this offset as a free parameter. We have yet to identify the physical origin of this phase offset, but find that it is small, $\phi_0=-0.09\pi$ and independent of $\tau$. A similar phase offset has been observed in other quantum acoustic experiments~\cite{cleland2023studying}. From this analysis we find that the dephasing rate for surface phonons in this device is $\kappa_{\phi}/2\pi~=~180\pm 20$~kHz.

While one might expect mechanical resonators to operate in a semi-classical regime free from dephasing, the loss of phase information in quantum acoustic resonators has previously been reported and has been attributed to interactions between the mechanical resonator and an ensemble of two-level fluctuators~\cite{wollack2022quantum, cleland2023studying} or frequency instability arising from thermal fluctuations of the qubit~\cite{gambetta2006qubit, chu2018creation, lane2020integrating} or time-varying magnetic fields. In this experiment, all of the above are plausible sources of dephasing. Both fluctuations in the thermal population of the qubit and its frequency are likely contributing to the SAW dephasing on some level. In the limit that $1/T_1 \gg \chi$, the SAW resonator dephasing due to thermal fluctuations in the qubit population is given by $\kappa_{\phi}^{th}=\chi^2T_1(1-\langle\hat{\sigma}_z\rangle^2)$~\cite{bachtold2022mesoscopic, dykmanchat}, where $\langle\hat{\sigma}_z\rangle$ is the expectation value of the qubit operator, which depends on the effective temperature of the system. $\kappa_{\phi}^{th}$ is maximized in the high temperature limit when $\langle\hat{\sigma}_z\rangle^2\rightarrow 0$. In this situation, $\text{max}(\kappa_{\phi}^{th})=\chi^2T_1$, which is significantly smaller than our experimentally measured value of $\kappa_{\phi}$ and thus SAW dephasing cannot be caused solely by thermal fluctuations in the qubit population. Additionally, interference between multiple surface phonon modes could manifest as dephasing. In fact, we have observed surface phonon-phonon interference in previous experiments with this device~\cite{kitzman2023fano}. While we are not able to determine the exact microscopic source of dephasing, our analysis based on the master equation indicates that loss of phase information contributes non-negligibly to the time evolution of the SAW resonant mode.

The energy decay and dephasing rates can be related to the overall decoherence rate $\kappa$, 

\noindent \begin{equation}
    \kappa = \kappa_1/2+\kappa_{\phi}.
\end{equation}
Using the energy decay and dephasing rates measured in the time domain, we calculate $\kappa~=~420\pm 25$~kHz. We can compare this to the absorption linewidth of the SAW resonator spectrum, $\kappa~=~430\pm27$~kHz, which we find is in good agreement. As with our measurements of the surface phonon energy loss, the methods we describe here for quantifying mechanical dephasing by comparing time and frequency-domain measurements can be applied more broadly across various quantum acoustic systems.

\section{\label{sec:conclusion}Conclusions}
In summary, we report on measurements and numerical modelling of decoherence in a SAW-based quantum acoustic device. The SAW resonator is driven into a coherent state by applying a microwave drive. Subsequent time-gated spectral measurements of a coupled transmon qubit are used to calculate the mean phononic population. In these time-resolved measurements we measure both energy decay and dephasing of the SAW resonant mode. We find excellent agreement between the data and modelling using the Markovian master equation, from which we find that the SAW resonant mode has considerable energy decay $\kappa_1/2\pi~=$~0.48~MHz and non-zero dephasing $\kappa_{\phi}/2\pi~=~0.18$~MHz, demonstrating the need to model the device as an open quantum system. 

The experimental benchmarking and analysis techniques described here can be applied in conjunction with iterative device fabrication to systematically control and understand the mechanisms for loss and dephasing in quantum acoustic systems in general, like those described in Ref.~\cite{gruenke2024surface}, where only frequency domain measurements were performed. Comparing frequency domain measurements with time-resolved measurements, as we do here, is important for circumventing potential complications due to interference~\cite{rieger2023fano}. Additionally, in the time-domain energy decay and dephasing processes can be disentangled from one another. Furthermore, as two-level fluctuators are a common noise source in quantum acoustic systems and devices, one could investigate methods to purposefully embed materials with fluctuating two-level systems in a systematic way to better understand their interaction with mechanical resonators. Future measurements could be performed to simultaneously measure qubit and mechanical resonator noise. Correlations between these measurements, or lack thereof, could reveal the nature of the noise sources affecting cQAD devices. 

\begin{acknowledgments}
We thank M.I.~Dykman, K.W.~Murch, W.P.~Halperin, N.R.~Beysengulov, A.J.~Schleusner, J.R.~Lane, J.~Zhang and P.K.~Rath for valuable discussions. We also thank R.~Loloee and B.~Bi for technical assistance and use of the W.M.~Keck Microfabrication Facility at MSU. This work was supported by the National Science Foundation (NSF) via grant number ECCS-2142846 (CAREER) and the Cowen Family Endowment.
\end{acknowledgments}

\appendix

\bibliographystyle{apsrev4-2}
\bibliography{apssamp}

\end{document}